\documentstyle[aps,pre]{revtex}

\begin{document}

\title {\bf Trapping with biased diffusion species}

\author{Alejandro D. S\'anchez \thanks{Fellow of CNEA, Argentina;
e-mail: sanchez@cab.cnea.gov.ar}}
\address{Grupo de F\'{\i}sica Estad\'{\i}stica
\thanks{http://www.cab.cnea.gov.ar/CAB/invbasica/FisEstad/estadis.htm}\\
Centro At\'omico Bariloche (CNEA) and
Instituto Balseiro (UNC),\\
8400 San Carlos de Bariloche, Argentina}

\maketitle

\begin{abstract}
We analyze a trapping reaction with a single penetrable trap, in a
one dimensional lattice, where both species (particles and trap)
are mobile and have a drift velocity. We obtain the density as
seen from a reference system attached to the trap and from the
laboratory frame. In addition we study the nearest neighbor
distance to the trap. We exploit a stochastic model previously
developed, and compare the results with numerical simulations,
resulting in an excellent agreement.
\end{abstract}


\section{Introduction}

With the aim of obtaining a description of diffusion--limited reactions several
generalizations of the classical Smoluchowski model have been proposed \cite{rice}.
A stochastic model (SM) coming from the theory of nuclear reactors \cite{galanin,mig}
has been adapted to model these reactions \cite{wio1,tesis}.
This framework results to be very adequate to describe different situations,
making it possible to obtain results in excellent agreement with simulations
\cite{wio1,tesis,wio2,wio3,wio4,wio5,wio6,wio7,pvec,labo,seba}, particularly
when the model equation can be exactly solved.

The trapping reaction $A+B\to B$ with a single trap and a uniform
initial density of particles in a one--dimensional lattice is one
of the simplest cases describing a certain kind of chemical
reaction. In the present work we analyze this situation within the
framework of the SM, when both species are mobile and have a
drift.

Diffusion--controlled reactions in presence of a biasing field are
of great interest in many areas. Let us mention a few instances:
(a) Due to the gravity field, all diffusion processes that take
place over the earth are affected in some quantity. For example,
gravity is responsible for the drift of water molecules diffusing
underground \cite{wat1,wat2}. (b)In positron tomography
\cite{posi} the addition of a strong field perpendicular to the
surface leads to greater sample penetration. (c) In problems of
electromigration \cite{elec} the model discussed here could be
used to analyze the effect of a trapping centers on carrier
concentration.

A situation related to the presented here was studied in Ref.
\cite{sibona1}, where the case of a stationary and impenetrable
trap was analyzed. In the mentioned work a diffusion equation with
adequate boundary conditions is the base of calculations. We want
to remark that in this paper we model a standard trap, where the
particles can eventually pass from left to right of it (or vice
versa) without being trapped. As we will see, while calculations
where an impenetrable trap is involved are naturally treated with
a diffusion equation, standard trap problems are naturally treated
within the SM. This difference is not relevant for particles
without a bias in its diffusive motion, at least for the density
calculations.

The organization of the paper is as follows. In the next section we briefly describe
the model and discuss the differences between penetrable and impenetrable traps. In
Sec. III we calculate the density as seen from a reference system attached to
the (mobile) trap and discuss some symmetries. In the next section we make a similar
analysis for the standard density, i. e. from the laboratory reference system, and
calculate the number of particles absorbed by the trap. In Sec. V we present the
results for the nearest neighbor distance. Finally we draw our conclusions.

\section{The model}

The model considers two species
of particles, $A$ and $B$, both mobile and independent, and having a given
reaction probability when they meet.
Here we will study a one--dimensional trapping reaction in a system
of $A$ particles uniformly distributed and a
single trap $B$, with both species having a biased diffusive transport.

The model equation for the evolution for $N(x,t)$, the density of the
$A$ particles for a given realization of the
trap trajectory $\epsilon (t)$, is the following
\begin{equation}
\label{gala}
\frac {\partial} {\partial t}N(x,t)=D_A \frac{\partial ^2}{\partial x^2}
N(x,t)-v_A \frac{\partial}{\partial x} N(x,t)-\gamma \delta \left( x -
\epsilon (t)\right) N(x,t),
\end{equation}
where $\gamma $ is a constant measuring the reaction probability
($\gamma \to \infty$ for perfect reactions), $D_{A,B}$ is the diffusivity and $v_{A,B}$
characterizes the bias of the $A$ or $B$ particles. In general we
are interested in $n(x,t)=\langle N(x,t) \rangle$, the density averaged
over realizations of $\epsilon (t)$ in the laboratory frame, or the similar
(averaged) density $\tilde n(x,t)$ seen from a reference system attached to the trap.

As was pointed in Ref. \cite{wio2} the SM equations given in Ref. \cite{wio1} are valid
using the propagators corresponding to the particle transport process. In our case we
have for $A$ and $B$ particles
\begin{eqnarray}
G(x,t)&=&\frac{1}{\sqrt{4 \pi D_A t}} \exp \left[ -\frac{(x-v_A t)^2}{4 D_A t}\right]\\
W(x,t)&=&\frac{1}{\sqrt{4 \pi D_B t}} \exp \left[ -\frac{(x-v_B t)^2}{4 D_B t}\right]
\end{eqnarray}
respectively.

We will consider for a moment that the trap is fixed at the origin and the $A$ particles
are not biased. Then Eq. (\ref{gala}) (in this case
$N(x,t)=n(x,t)=\tilde n(x,t)$) becomes
\begin{equation}
\label{galaf}
\frac {\partial} {\partial t}n(x,t)=D_A \frac{\partial ^2}{\partial x^2}
n(x,t)-\gamma \delta \left( x \right) n(x,t).
\end{equation}
The resulting equation is equivalent to a diffusion equation except at the
origin. It is not difficult to obtain the jump of the derivative in the density
\cite{wio7}, which due to the symmetry results in the following boundary condition
(for $x>0$)
\begin{equation}
\label{bc} \frac{\partial}{\partial_x}
n(x,t)|_{x=0}=\frac{\gamma}{2 D_A} n(0,t).
\end{equation}
Hence the diffusion equation with this boundary condition is
equivalent to Eq. (\ref{galaf}). The same boundary condition could
be obtained requiring that the diffusive current at the trap
position (left minus right contributions) is proportional to the
density, which is the usual way to obtain the so called radiative
boundary condition for an imperfect trap. However, this is true
due to the symmetry, which requires that the current at both sides
of the trap are of equal magnitude. When coming to our bias
problem (but with the trap immobile) the particles have a
preferred direction of motion and the symmetry argument is not
valid. On one hand, Eq. (\ref{gala}) models a penetrable trap but,
although we can write an equation for the jump of the derivative
we can not to write a equation for each lateral derivative
separately. On the other hand, a current proportional to the
density (from one particular direction) models an impenetrable
trap. In this way we can see that each model is the natural way to
treat each different situation. It is intuitive that for perfect
trapping ($\gamma \to \infty$) both models give the same density.
A final remark is that the coefficient relating the diffusive
current and the density at the trap position was taken as
$\gamma$. This election make both models compatible. We must
change $\gamma$ by $\gamma/(2 D_A)$ in the equations of Ref.
\cite{sibona1} in order to make a direct comparison with results
given in this work.

A detailed discussion for unbiased particles (and fixed trap) of penetrable and
impenetrable traps was done by Taitelbaum \cite{taite0}.
Note that the analysis done here about impenetrable traps is only for a fixed trap,
the case considered in Ref. \cite{sibona1}. The treatment for a mobile impenetrable
trap is a non trivial generalization, and is out of the scope of this work.

\section{Density in the trap frame}

In complete analogy with the case without bias \cite{pvec} we change the reference
system to one fixed to the trap. After doing the average it results that the density
is similar to the case with a fixed trap, but changing $D_A$ by $D_1=D_A+D_B$ and
$v_A$ by $v_r = v_A-v_B$. The resulting equation is solved by Laplace transform
techniques, yielding after some calculations the expression
\begin{eqnarray}
\label{enie}
\frac{\tilde n(x,t)}{n_0}&=&1-\frac{1}{2} \frac{{\rm e}^{v_r (x-|x|)/(2 D_1)}}
{1+v_r / \gamma} {\rm erfc} \left( \frac{|x|-v_r t}{\sqrt{4 D_1 t}}\right)
-\frac{1}{2} \frac{{\rm e}^{v_r (x+|x|)/(2 D_1)}}
{1-v_r / \gamma} {\rm erfc} \left( \frac{|x|+v_r t}{\sqrt{4 D_1 t}}\right)\\ \nonumber
&&+\frac{1}{1-(v_r/\gamma)^2}{\rm e}^{(v_r x+\gamma |x|)/(2 D_1)+(\gamma^2-v_r^2)t/(4 D_1)}
{\rm erfc} \left( \frac{|x|+\gamma t}{\sqrt{4 D_1 t}}\right),
\end{eqnarray}
valid for any value of the parameters except if $|v_r|=\gamma$. For $v_r=\gamma$ the
expression is (expression for $v_r=-\gamma$ can be obtained performing a specular
reflection)
\begin{eqnarray}
\frac{\tilde n(x,t)}{n_0}&=&1-\frac{1}{4} {\rm e}^{v_r (x-|x|)/(2 D_1)}
{\rm erfc} \left( \frac{|x|-v_r t}{\sqrt{4 D_1 t}}\right)
+\left[ \frac{v_r}{4 D_1}(|x|+v_r t)+\frac{1}{4} \right] \\ \nonumber
&& \times
{\rm e}^{v_r (x+|x|)/(2 D_1)}{\rm erfc} \left( \frac{|x|+v_r t}{\sqrt{4 D_1 t}}
\right)-\frac{v_r}{2}\sqrt{\frac{t}{\pi D_1}} {\rm e}^{-(x - v_r t)^2/(4 D_1 t)}.
\end{eqnarray}
We can see that $\tilde n$ depends on the drift velocities only through the relative
velocity $v_r$, and when both species have the same drift we recover the unbiased
case. \cite{labo,lab} In addition, a change in the sign of $v_r$ produce a specular
reflection in the density profile. Hence, without loss of generality in the analysis
we often set $v_r>0$. Note that in Eq. (\ref{enie}) the variables and parameters come
only in three combinations: $x/\sqrt{D_1 t}$, $v_r \sqrt{t/D_1}$ and $\gamma /v_r$.

Figure 1 shows a typical $\tilde n$ at different times and four sets of simulations
corresponding to the same theoretical profile as all have the same $D_1$ and
$v_r$. The agreement between the SM result and simulations is excellent.

We can observe that for $t \to \infty$ the density in the trap position approaches
\begin{equation}
\label{enie0}
\frac{\tilde n(0,t)}{n_0} \sim \frac{|v_r|}{\gamma+|v_r|},
\end{equation}
which is a positive value (for imperfect trapping), in contrast with the unbiased case
($v_r=0$) where the particle density asymptotically reaches zero at the trap
position.\cite{labo,lab}
Besides, for $v_r>0$ the density reaches a stationary profile
($\tilde n(x,t)/n_0 \sim 1-\exp (v_r x /D_1)/(1+v_r/\gamma)$) in the left semi--axis,
i. e. where the relative drift is towards the trap, while any point to the right of
the trap reaches asymptotically the value given by Eq. (\ref{enie0}). These aspects
can be seen in Fig. 1.

\section{Density in the laboratory frame}

The Laplace--Fourier transform of the density in the laboratory frame \cite{wio1}
is given by
\begin{eqnarray}
\label{nks}
\frac{n(k,s)}{n_0}&=&\frac{2 \pi \delta(k)}{s}-\frac{\gamma}
{\left(s+D_A k^2+i k v_A \right) \left(s+D_B k^2 + i k v_B \right)} \nonumber \\ & &
\times \frac {1}{1+\gamma \left[4 D_1 s+ v_r^2+
4 D_A D_B k^2+4 i k (v_A D_B+v_B D_A )\right]^{-1/2}}.
\end{eqnarray}
Here we can see that Eq. (\ref{nks}) does not change if we interchange $D_A$ by $D_B$
and $v_A$ by $v_B$ simultaneously. This fact shows that the trap and the particles play exactly
the same role in the density. However, this characteristic is not valid for any magnitude,
as is known from the first neighbor distance in the unbiased problem \cite{pvec}.
The equivalence between both particle species just mentioned lead to two interesting
consequences that we describe in the following.

The first consequence is a symmetric density
if we set $D_A=D_B=D$ and $v_A=-v_B=v$. This can be verified considering that from
Eq. (\ref{nks}) results $n(k,s)=n(-k,s)$. To obtain the general expression in the $x$
and $t$ variables is very difficult and we present the simple result for perfect
absorption
\begin{eqnarray}
\frac{n(x,t)}{n_0}&=&\frac 1 2+\frac 1 2 {\rm erf} \left( \frac{x+vt}
{\sqrt{4 D t}} \right){\rm erf} \left( \frac{x-vt}{\sqrt{4 D t}} \right)\nonumber \\
&&+\frac{2}{v} \sqrt{\frac{D}{\pi t}} \left [  {\rm e}^{-(x-v t)^2/(4 D t)}
{\rm erf} \left( \frac{x+vt}{\sqrt{4 D t}} \right)- {\rm e}^{-(x+v t)^2/(4 D t)}
{\rm erf} \left( \frac{x-vt}{\sqrt{4 D t}} \right) \right ].
\end{eqnarray}
This expression has only two variables (any parameter) that are $x/\sqrt{4 D t}$
and $v \sqrt{t/(4 D)}$. We note that this expression approaches asymptotically
zero at the origin, in contrast with the unbiased case where the density reaches a
finite value. \cite{labo,lab} In Fig. 2 we can see an example of symmetric profiles at
different times. The agreement between theory and simulation is good, although there
is a systematic difference. However, note that the symmetry is still present in
simulations.

The second consequence is that setting $D_B=v_B=0$ (fixed trap) the density expression is
the same as setting $D_A=v_A=0$ (fixed particles) exchanging the particle subindexes.
Moreover, since
for the fixed trap we have the trivial equality $\tilde n(x,t)=n(x,t)$, and
$\tilde n(x,t)$ for $D_B=v_B=0$ is equal to $\tilde n(-x,t)$ for $D_A=v_A=0$,
we have $\tilde n(-x,t)=n(x,t)$ for fixed particles, which is not evident.
A similar property was found in the unbiased case \cite{labo}. In Fig. 3 we
show simulations of $\tilde n(x,t)$ and $n(x,t)$ for fixed particles, $n(x,t)$
for the fixed trap and the SM result corresponding to all of them. We can see an
excellent agreement.

By setting $k=0$ in the second term of Eq. (\ref{nks}), the total number of
absorbed particles in the Laplace domain can be obtained. Performing the inverse
transform we obtain
\begin{eqnarray}
\label{nabs}
\frac{N_{ABS}(t)}{n_0}&=&\frac{2 \gamma^2 \sqrt{D_1 t}}{\sqrt{\pi}(\gamma^2-v_r^2)}
{\rm e}^{-v_r^2 t/(4 D_1)}+\frac{4 D_1 \gamma^3}{(\gamma^2-v_r^2)^2}
{\rm e}^{(\gamma^2-v_r^2)t/(4 D_1)}{\rm erfc}\left( \gamma \sqrt{\frac{t}{4 D_1}}\right)
\nonumber \\ &&+\frac{\gamma |v_r| t}{\gamma+|v_r|}+\frac{2 D_1}{|v_r|}
\frac{\gamma^2}{(\gamma+|v_r|)^2} \nonumber \\ &&-\left[ \frac{2 D_1 \gamma^2}{|v_r|}
\frac{\gamma^2+v_r^2}{(\gamma^2-v_r^2)^2}+\frac{\gamma^2 |v_r| t}{\gamma^2-v_r^2} \right]
{\rm erfc}\left( |v_r| \sqrt{\frac{t}{4 D_1}}\right).
\end{eqnarray}
The limit cases $\gamma \to \infty$ and $\gamma=|v_r|$, not included for the
sake of brevity, can be extracted from Eq. (\ref{nabs}) taking the corresponding limit.
In Fig. 4 we show simulations of $N_{ABS}$ for the same sets of parameters than in
Fig. 1. Simulation data coming from all sets are rather coincident (almost
indistinguishable in the figure) and in excellent agreement with the SM prediction.
If $v_A=v_B$ Eq. (\ref{nabs}) coincides with the case without bias, where
asymptotically $N_{ABS} \propto t^{1/2}$\cite{pvec}. However for $v_r=v_A-v_B \neq 0$
the asymptotic behavior results
\begin{equation}
\frac{N_{ABS}(t)}{ n_0} \sim \frac{\gamma |v_r|}{|v_r|+\gamma} t.
\end{equation}
This change in the behavior can be understood from Eq. (\ref{enie0}), since the
density at the trap position reaches a positive value. Hence, the absorption rate
$\gamma \tilde n(0,t)$ becomes constant. The same arguments are applicable
to an impenetrable trap that has the same qualitative dependence (although the factor
is different because $\tilde n(0,t)$ is different).

It is possible to factorize Eq. (\ref{nks}) (after performing the inverse Laplace
transform)
in $\exp(-i k v_B t)$ and an expression where $v_A$ and $v_B$ arise only in the $v_r$
combination. This implies that in the $x$ domain the density has the form
$n(x,t)=n^*(x-v_B t)$, where $n^*$ depends on $v_r$ but neither on $v_A$ nor $v_B$
separately. Due to the identical role played in the density expression by both
particle species, we can do a similar factorization involving $\exp(-i k v_A t)$.

In Eq. (\ref{nks}) we take the limit $\gamma \to \infty$ in order to obtain
the mathematically simple expression for perfect reaction resulting, after performing the
inverse Laplace transform, in
\begin{eqnarray}
\label{nk}
\frac{n(k,t)}{n_0}&=&2 \pi \delta(k)-\frac{2}{k^2(D_B-D_A)-i k v_r} \left \{
(i k D_A-\frac{v_r}{2}) {\rm e}^{-(D_A k^2+i k v_A)t} {\rm erf} \left[
(i k D_A-\frac{v_r}{2}) \sqrt{\frac{t}{D_1}}\right]  \right . \nonumber \\
 && \left . -
(i k D_B+\frac{v_r}{2}) {\rm e}^{-(D_B k^2+i k v_B)t} {\rm erf} \left[
(i k D_B+\frac{v_r}{2}) \sqrt{\frac{t}{D_1}} \right ] \right \}.
\end{eqnarray}
Numerically performing the inverse Fourier transform we obtain the density in the $x$
domain. In Fig. 5 we
show numerical integrations at different times together with the respective
simulations. The agreement between both results is apparent.

\section{Nearest neighbor--distance}

As was stated in Ref. \cite{pvec}, we can use the expression for the PDF calculated
in a similar way as for fixed particles as an approximation to the problem of both
mobile species. This approximation is better for a small $D_B/D_A$ ratio and for
short times. We note that we can consider  biased particles as having a deterministic drift
velocity plus an unbiased diffusion. Hence, following Ref. \cite{pvec} the PDF only
depends on $v_r$ (at variance with diffusivities where the dependence is not trivial).
To calculate the PDF of the (one sided) nearest neighbor distance we first compute
the intermediate
quantity $Q(x,t)=\exp \left( -\int_0^x \tilde n(u,t) du \right )$, and then the PDF is
$p(x,t)=-\partial_x Q(x,t)=\tilde n(x,t) Q(x,t)$.\cite{weiss,redner,imper} The
integral involved in $Q$ can be
done analytically, yielding a rather complicated expression.

In Fig. 6 we show a plot of $p$ corresponding to the right neighbor and compare it with simulations.
We can see a good agreement. The most noticeable difference arises near
the origin. However, the mean distance $\langle x \rangle=\int_0^\infty p(x,t)x
dx=\int_0^\infty Q(x,t) dx$ shows the expected systematic differences due to the
approximation.
This is apparent in Fig. 7 where we show $\langle x \rangle$ coming from a numerical
integration.
Note that squares and diamonds are almost coincident in Fig. 7 (because they
have the same parameters except the drift velocities but with the same $v_r$).
In addition we make a similar calculation and simulations for the left neighbor.
In this case the mean distance reaches a stationary value that can be calculated within the
SM and is
\begin{equation}
\langle x \rangle=\frac{D_1}{v_r}\left[ \left( 1+\frac{v_r}{\gamma}\right)
\frac{v_r}{n_0 D_1} \right ]^{n_0 D_1/v_r} \exp \left ( \frac{n_0 D_1}{v_r}
\frac{\gamma}{\gamma+v_r} \right)
\overline{\gamma} \left( \frac{n_0 D_1}{v_r},\frac{n_0 D_1}{v_r}
\frac{\gamma}{\gamma+v_r} \right),
\end{equation}
where $\overline{\gamma}(a,z)$ is the incomplete gamma function \cite{abra} (we have
the over bar to avoid confusions with the absorption constant). We can see in the figure
that this quantity does not have a good agreement with simulations. This is due to the
fact that the model is a continuous one and can not describe well the situation very
near the trap, where the discreteness of the problem is relevant. This phenomenon
is the same as in Fig. 6 (small $x$), but here it is more noticeable because almost all
particles in the simulation are in the same place of the trap or in the nearest
neighbor site.

Simulations were performed on a lattice of $L$ sites, with periodic boundary
conditions.
We have used the same algorithm described in \cite{pvec}, with a small change to
allow for the drift of the particles. The connections between the simulation and
SM parameters
are $D=\omega_j \Delta x^2/2$, $v=(2 q-1)\omega_j \Delta x$ and $\gamma=\omega_r
\Delta x$; where $\Delta x$ is the jump length, $\omega_{j,r}$ are the jump and
reaction frequencies respectively, and $q$ is the probability of a given jump
to the right ($q=0.5$ corresponds to an unbiased particle).
All show simulations are the average of $10 \; 000$ realizations over a 200 site
lattice and with an initial density $n_0=1$.

\section{Conclusions}

We have studied a trapping reaction $A+B \to B$ with a single penetrable trap $B$
in a one dimensional lattice,
where both species are mobile and perform a biased diffusion.

We have seen that the natural tool to perform analytical calculations is the stochastic model
(SM) in contrast
with a diffusion equation that is useful for calculations involving an impenetrable trap.
We have calculated the expressions of $\tilde n(x,t)$, the distribution of particles in
the trap frame, and $n(x,t)$, the usual density in the laboratory frame.

Analogously to the unbiased case, $\tilde n(x,t)$ can be obtained from the fixed trap case
doing the exchange $D_A \to D_1=D_A+D_B$ and $v_A \to v_r=v_A-v_B$
in the corresponding expression. In the general expression we note that $\tilde n(0,t)$
for imperfect reactions reaches a non zero value at long time. This results in a linear
asymptotic number of absorbed particles, in contrast with the $t^{1/2}$ behavior of
the unbiased problem.

The expression for $n(x,t)$ has the following symmetry: it is
invariant by exchange of diffusivities and velocities
simultaneously, i. e. $D_A \to D_B$, $D_B\to D_A$, $v_A \to v_B$,
$v_B\to v_A$. This is valid both for perfect and imperfect
reactions. Due to this characteristic there is a combination of
parameters, namely $D_A=D_B$ and $v_A=-v_B$, which results in a
symmetric density. Another consequence is that $n(x,t)=\tilde
n(-x,t)$ for fixed $A$ particles, a property found first in the
unbiased case. We note that in the symmetric case for perfect
reaction the particles are asymptotically extinguished, in
contrast with the unbiased case where the density reaches a
positive value. An explanation of this fact is that the trap in
average goes to the right and particles located to its right can
not pass through it to reach the origin (because it is a perfect
trap). In addition particles situated to the left have in average
a velocity to the left. Hence the origin results asymptotically
free of particles yielding the observed null density.

A comparison of the density profile with that obtained for an impenetrable trap,
when the trap is fixed (where we have available results from Ref. \cite{sibona1}),
shows that both expressions are similar if the absorption rate is high. In fact,
for perfect absorption both densities become identical. On the contrary, for low
absorption an impenetrable trap implies that the untrapped particles are
accumulated (to the left if
$v_A>0$) while they follow their way to the right for a penetrable trap. In spite of
this, both problems reach a stationary profile to the left of the trap (for $v_A>0$).

We have studied the nearest neighbor distance in the case of $D_B=0$, and used an
approximation valid for small $D_B$. We can see that the mean distance, after a short
transient, approaches a constant to the left of the trap, while it grows to the right
of the trap.

Simulations are in excellent agreement with the model. As was
mentioned in the introduction, this is an expected result since
the model has been solved exactly. Finally, we want to remark that
while in unbiased particle problems the SM is a convenient
alternative to other technics, in problems of biased particles it
becomes the natural way to obtain analytical results.

\acknowledgments
Financial support from CONICET (Project PIP-4953/96) and ANPCyT (Project
03--00000--00988), Argentina is acknowledged. The author wants to thank
H. S. Wio for the careful revision of the manuscript.

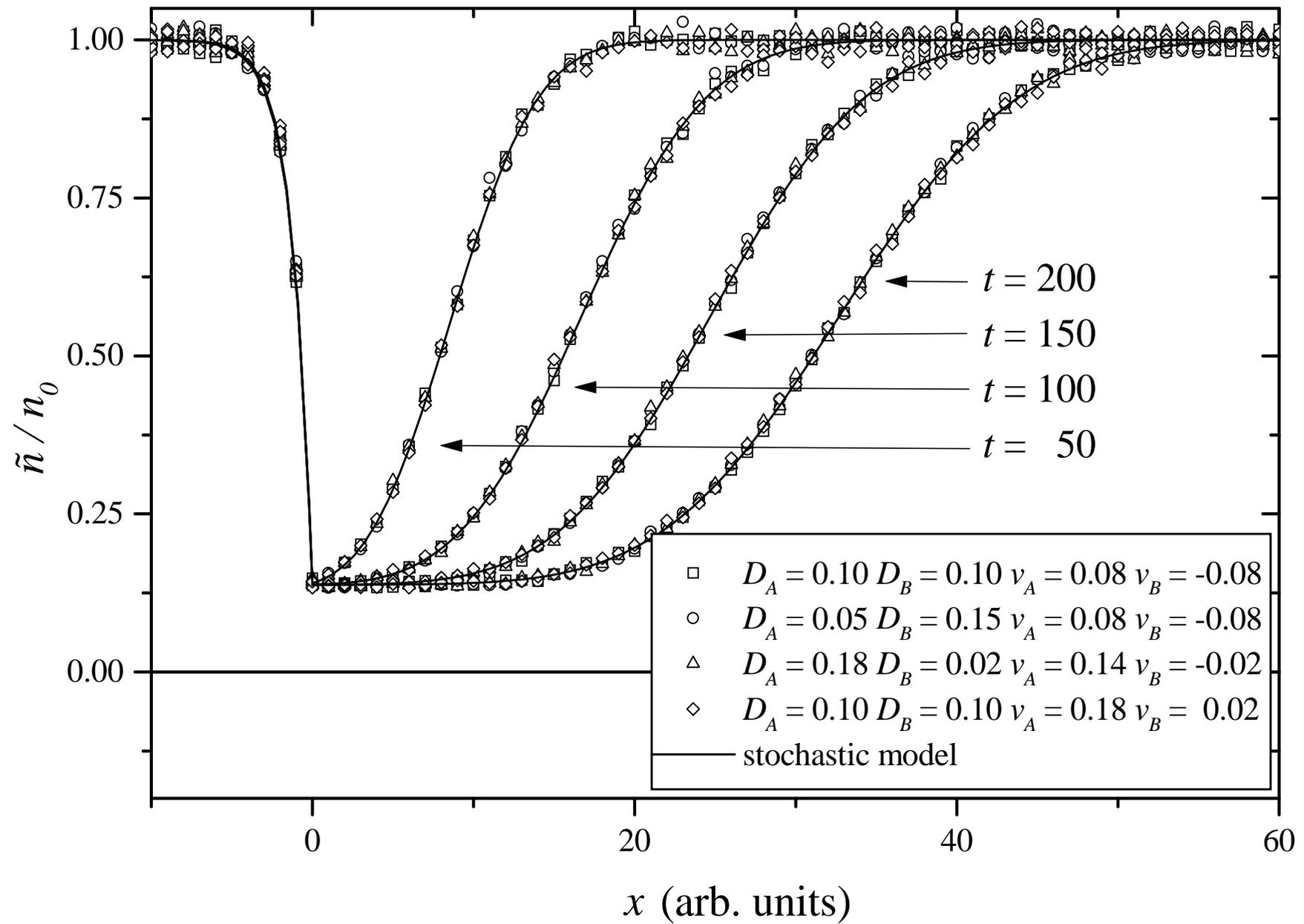
\begin{figure}
\caption{Density of $A$ particles in the reference frame of the trap at different
times. Simulations
are performed with different parameter sets indicated in the figure. All of them
have $D_1=0.2$ and $v_r=0.16$. The reaction is imperfect with $\gamma=1$.}
\end{figure}

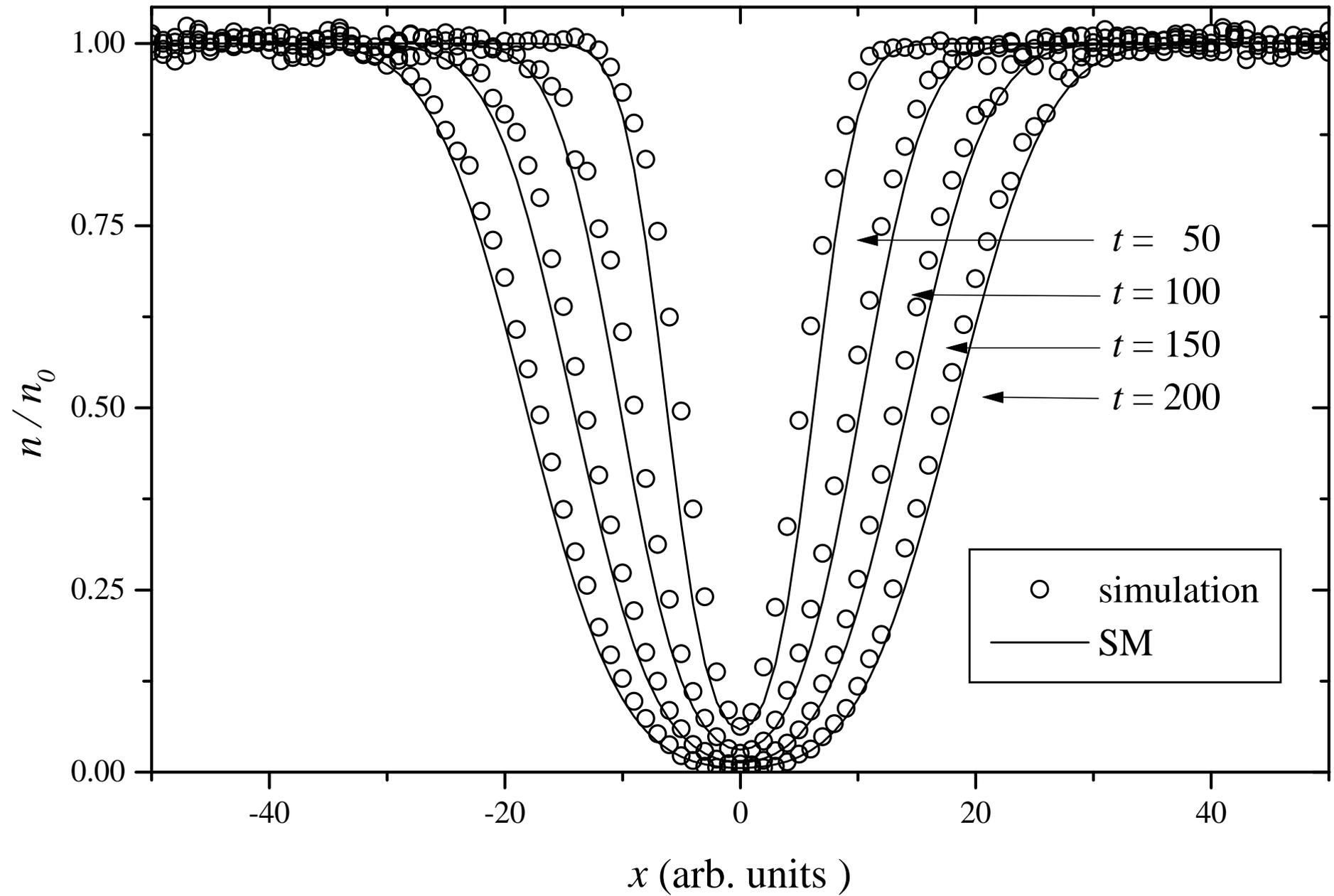
\begin{figure}
\caption{Density of $A$ particles in the symmetric case, as seen from the laboratory
coordinate system, for a perfect reaction. The simulation parameters are:
$D_A=D_B=0.1$ and $v_A=-v_B=0.08$.}
\end{figure}

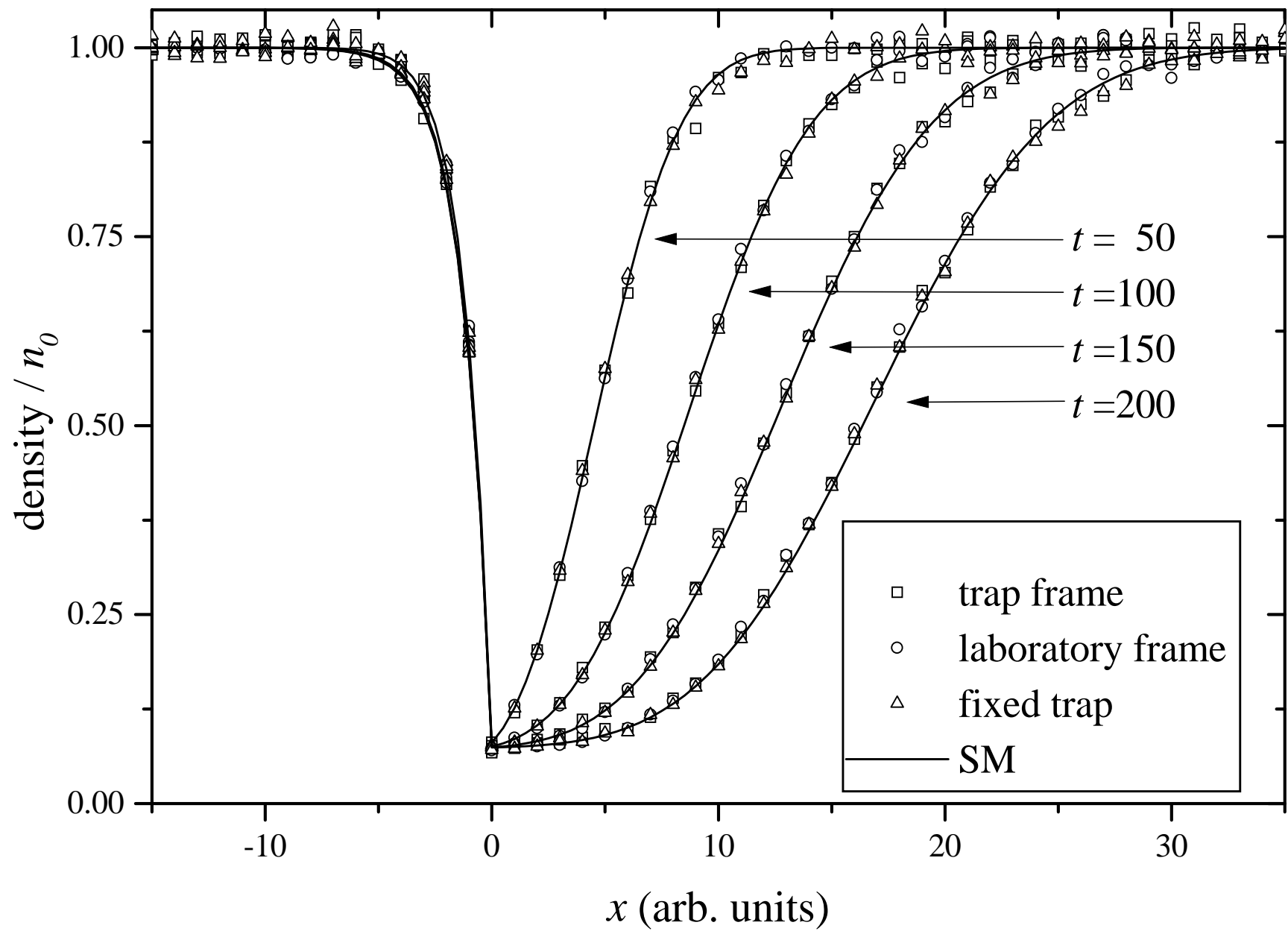
\begin{figure}
\caption{Simulation for the density of fixed $A$ particles in the
reference frame of the trap, from the laboratory coordinate system, and for the fixed
trap. The parameters are:
$D_A$ or $D_B=0.1$, $|v_A|$ or $|v_B|=0.08$ and $\gamma=1$.}
\end{figure}

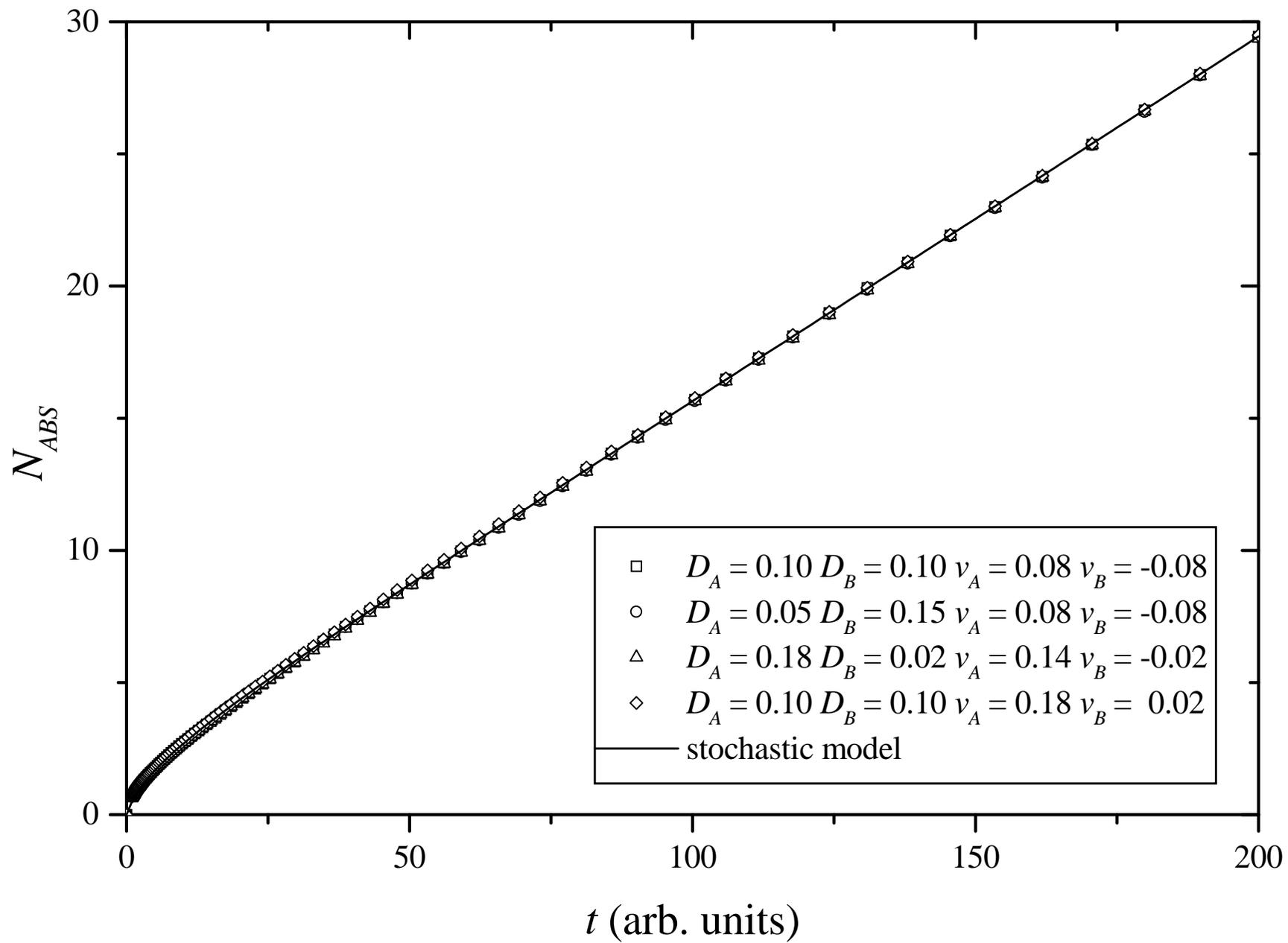
\begin{figure}
\caption{Number of absorbed particles. The parameters are the same as
those of the Fig. 1}
\end{figure}

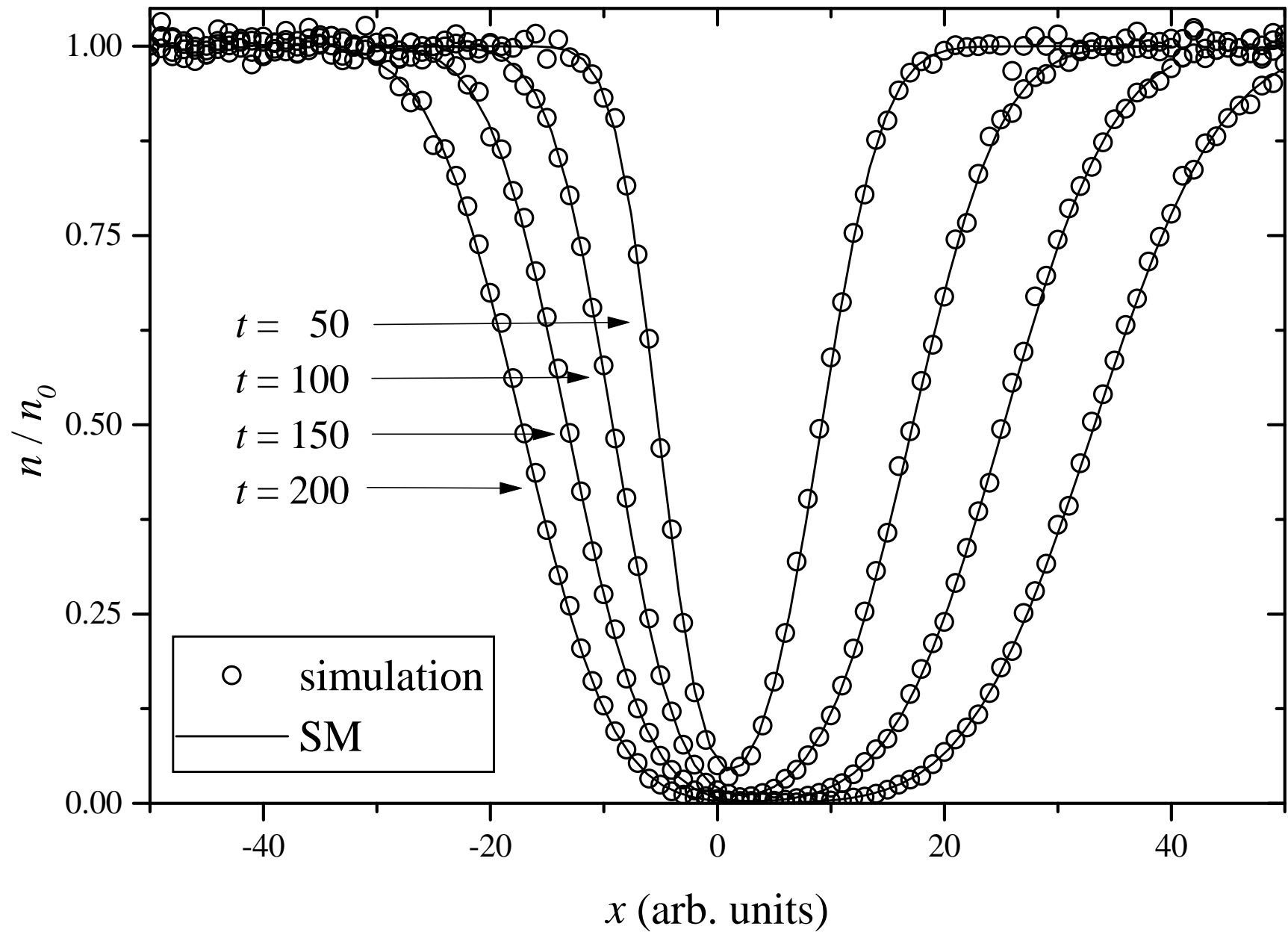
\begin{figure}
\caption{Density of $A$ particles, as seen from the laboratory coordinate system,
for a perfect reaction. The simulation parameters are: $D_A=0.2$, $D_B=0.1$,
$v_A=0.16$ and $v_B=-0.08$.}
\end{figure}

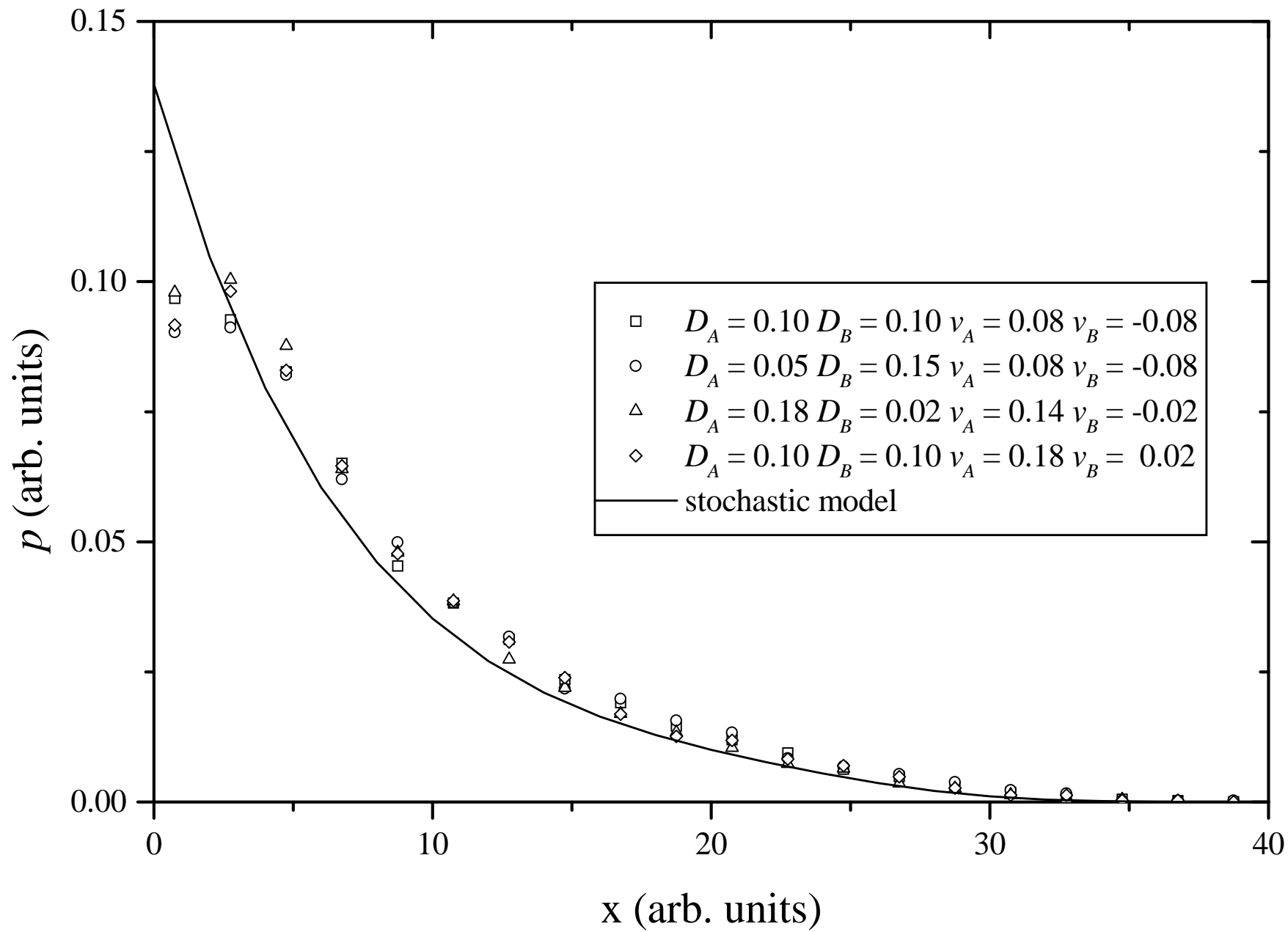
\begin{figure}
\caption{PDF of the nearest neighbor distance to the trap at $t=200$. The parameters,
indicated in the figure, are the same as in Fig. 1.}
\end{figure}

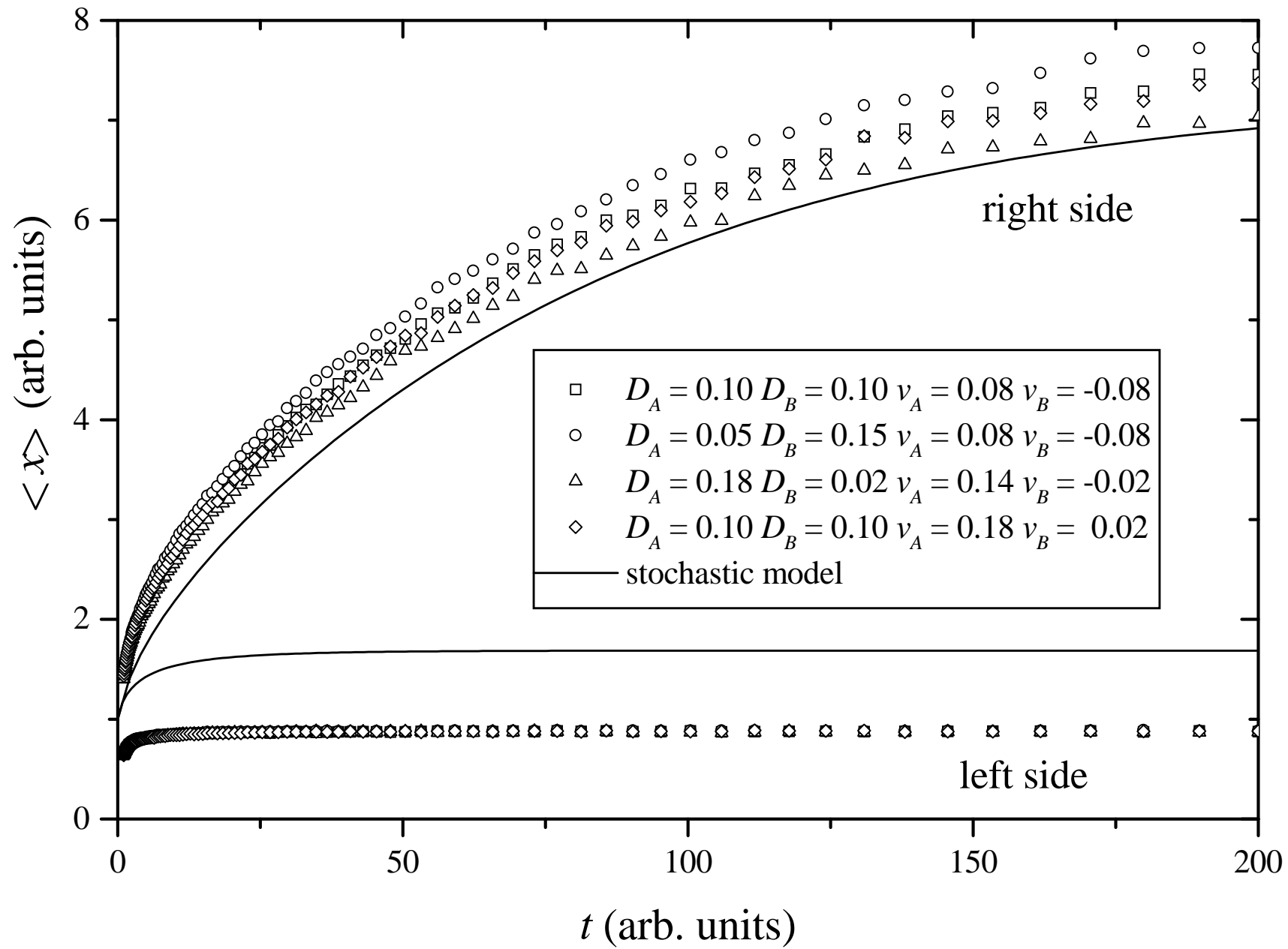
\begin{figure}
\caption{Mean distance to the nearest neighbor to the trap. The parameters,
indicated in the figure, are the same as in Fig. 1.}
\end{figure}

\end{document}